\documentclass[%
reprint,
twocolumn,
superscriptaddress,
showpacs,preprintnumbers,
 amsmath,amssymb,
pra,
]{revtex4-1}

\usepackage{graphicx}
\usepackage{dcolumn}
\usepackage{bm}
\usepackage{color}

\def\Put(#1,#2)#3{\leavevmode\makebox(0,0){\put(#1,#2){#3}}}

\begin{document}


\title{All-optical transport and compression of ytterbium atoms into the surface of a solid immersion lens}

\author{M. Miranda}
\email{miranda.m.aa@m.titech.ac.jp}
\author{A. Nakamoto}
\author{Y. Okuyama} 
\author{A. Noguchi}
\altaffiliation[Current affiliation: ]{Osaka University}
\affiliation{Department of Physics, Tokyo Institute of Technology,\\2-12-1 O-okayama, Meguro-ku, Tokyo 152-8550, Japan}
\author{M. Ueda}
\affiliation{Department of Physics, University of Tokyo,\\Hongo, Bunkyo-ku, Tokyo 113-0033, Japan}
\author{M. Kozuma}
\affiliation{Department of Physics, Tokyo Institute of Technology,\\2-12-1 O-okayama, Meguro-ku, Tokyo 152-8550, Japan}

\date{\today}

\begin{abstract}
We present an all-optical method to load ${}^{174}$Yb atoms into a single layer of an optical trap near the surface of a solid immersion lens which improves the numerical aperture of a microscope system. Atoms are transported to a region 20 $\mu$m below the surface using a system comprised by three optical dipole traps. The ``optical accordion'' technique is used to create a condensate and compress the atoms to a width of 120 nm and a distance of 1.8 $\mu$m away from the surface. Moreover, we are able to verify that after compression the condensate behaves as a two-dimensional quantum gas.
\end{abstract}

\pacs{37.10.De, 37.10.Gh, 67.85.Hj, 67.85.Jk}
\maketitle

\section{\label{sec:level1} Introduction}

Neutral atoms trapped in a two-dimensional optical lattice have been demonstrated to be a novel candidate for studying interacting many-body quantum systems \cite{PhysRevA.80.021602, PhysRevLett.92.173003} and creating quantum simulators \cite{springerlink:10.1007/BF02650179, nature09994}. One of the most challenging aspects in the study of two-dimensional optical lattice systems was how to measure and manipulate the atoms in each site of the lattice. In recent years, several techniques have been developed to observe the atoms with the help of high numerical aperture microscope systems \cite{nature08482, nature09378}. In particular, a system consisting of a solid immersion lens (SIL) and a high numerical aperture objective lens was utilized to detect single rubidium atoms in a two-dimensional optical lattice \cite{nature08482}. By simply introducing the solid immersion lens between the sample and the objective lens, the resolution of the system can be improved by a factor of the index of refraction of the hemispherical lens \cite{mansfield:2615}.

For a variety of quantum information processing, mainly represented by a quantum computer, it is required that the system is scalable, with the possibility of single-site manipulation and long coherence time for the quantum bit. The high scalability of a two-dimensional optical system, combined with a high resolution microscope system is then an ideal candidate for a quantum computer. To obtain a long coherence time, alkali-earth like atoms such as Ca, Sr, Yb and Hg are preferable, since they have no electronic spin, which ensures a long coherence time compared with alkali atoms \cite{PhysRevLett.101.170504}. In particular,  ${}^{171}$Yb has a 1/2 nuclear spin which is ideal for implementing a quantum bit \cite{springerlink:10.1007/s00340-009-3696-4, PhysRevA.84.030301, PhysRevA.81.062308}. In addition, most of the optical transitions in Yb are in the visible range, which ensures a high quantum efficiency and a high resolution for the microscope system.

Observation by a high resolution microscope system requires that atoms remain confined in a pancake-shaped region which is thinner than the depth of field of the objective lens. So far, several methods have been realized to create a two-dimensional system \cite{0953-4075-38-3-007, nature09378, PhysRevA.80.021602}. One of the methods consists of loading the atoms in a one-dimensional lattice, and then selecting a single slice from it by using a gradient magnetic field \cite{nature09378}. Another method is to compress an atomic cloud between the gradient potential of a magnetic field and a repulsive potential created by a blue-shifted evanescent wave \cite{PhysRevA.80.021602}. Both of these approaches are not applicable to atoms with an extremely small magnetic moment such as Yb.

In this paper we report an all-optical approach to load ultracold atoms into a single layer of a standing wave near the surface of solid immersion lens. This method can be extended to other experiments using species with low magnetic moment as Sr, Ca and Hg.  As the fermionic isotope ${}^{171}$Yb has a very small scattering length \cite{JPSJ.78.012001} making it unsuitable for evaporative cooling, we start our experiments with the ${}^{174}$Yb bosonic isotope. A fermi degenerate gas of ${}^{171}$Yb can be subsequently formed by sympathetic cooling by using  ${}^{174}$Yb as a coolant \cite{JPSJ.78.012001}.

\section{\label{sec:level2} Experimental setup and transport of atoms}

\begin{figure}[!ht]
\includegraphics{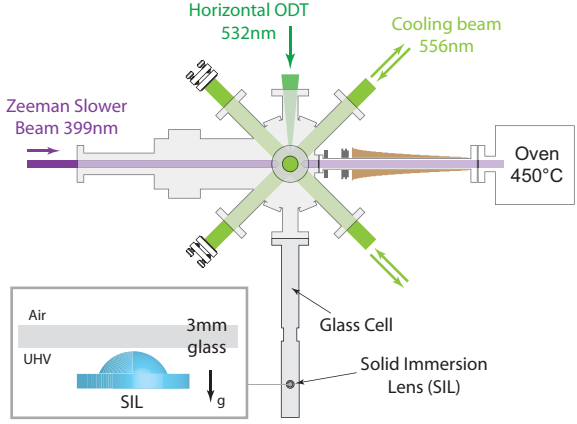}
\caption{Top view of the ultra-high-vacuum chamber. Atoms are slowed down by an increasing Zeeman slower, and then trapped with a 3D magneto optical trap. Attached to the metal chamber is a glass cell with 3mm thickness containing the SIL.}
\label{figure0}
\end{figure}

Our experimental setup consists of an ultra-high-vacuum (UHV) chamber in which the ultracold Yb atoms are prepared by using a magneto optical trap (MOT), and a glass cell containing the SIL (figure \ref{figure0}). The spatial separation between the chamber and the SIL prevents the atomic beam from contaminating the surface of the lens, and allows a high optical access to the atoms.

The first step of our experiment is to prepare ultracold ${}^{174}\text{Yb}$ atoms with a MOT \cite{PhysRevA.60.R745}. Atoms are slowed down with an increasing Zeeman slower and a laser beam detuned by  750 MHz to the ${}^1S_0 - {}^1P_1$ transition ($\Gamma/2\pi = 29$ MHz, $\lambda = 399$ nm). The cooling beam for the MOT is in a 3D retro-reflective configuration and is tunned to the narrow intercombination line $^1S_0 - {}^3P_1$ ($\Gamma/2\pi = 182$ KHz, $\lambda = 556$ nm). After 12 seconds of loading time, we obtain $8 \times 10^8$ atoms at a temperature of 27 $\mu$K. The lifetime of the MOT was approximately 3 minutes. To transport the cold atoms into the glass cell, we load the atoms in a horizontal optical dipole trap (ODT) with 15W of power at 532nm and a beam waist of 23$\mu$m. The resultant potential has a depth of 800$\mu$K. To effectively load the atoms from the MOT to the optical trap, we gradually compress the atoms by increasing the magnetic field, while lowering the intensity and decreasing the detuning of the cooling beam. Approximately $2 \times 10^7$ atoms are loaded into the horizontal ODT. Atoms are transported by mechanically moving the focusing lens system that produces the ODT along the light propagating axis. In order to avoid heating due to mechanical vibrations, the lens system is placed over an air-bearing linear stage (Aerotech ABL1500). The time of transportation is 2 seconds for a distance of 450mm, with the final number of $6 \times 10^6$ atoms.

\begin{figure}[!ht]
\includegraphics{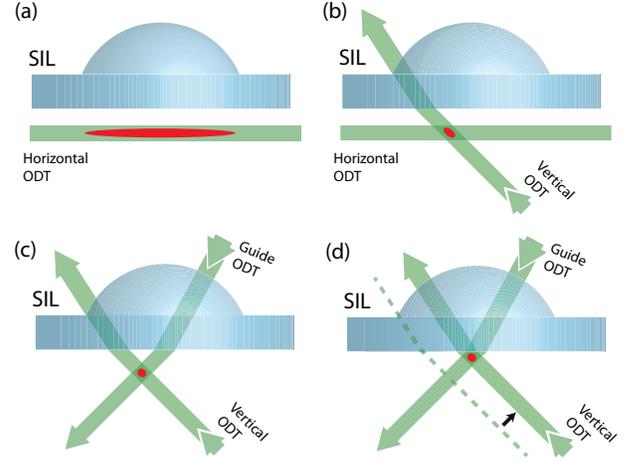}
\caption{Experimental setup for evaporative cooling and transport of ${}^{174}$Yb atoms to the surface of the SIL. a) Atoms are transported using a horizontal ODT to a distance of 300 $\mu$m under the SIL. b) Using a cross trap, atoms are cooled down to 2$\mu$K. c) Atoms are transfered to the guide ODT and the horizontal beam is removed. d) The vertical ODT is displaced and the atoms move along the guide ODT to the center of the solid immersion lens.}
\label{figure1}
\end{figure} 

After transportation, the atoms are positioned at a distance of 300 $\mu$m from the surface of the SIL as seen in figure \ref{figure1}(a). The atomic cloud has the dimensions of 1 mm and 25 $\mu$m in the axial and radial directions respectively, and a temperature of 47 $\mu$K. At this point we perform a forced evaporative cooling by crossing a vertical beam and reducing the power of the horizontal ODT in an exponential manner. The vertical ODT crosses the horizontal ODT in such an angle that the incident angle of the vertical ODT into the SIL is at the Brewster angle as shown in figure \ref{figure1}(b). We choose this angle to avoid reflections of the vertical ODT from the plane surface of the SIL, that would generate a standing wave in the vicinity of the surface, interfering with the transportation of the atoms. The polarization was also selected using a Glan-Taylor prism, effectively reducing the reflection by 5 orders of magnitude. Forced evaporation is performed in 2 seconds, and $8 \times 10^5$ atoms at 2$\mu$K are obtained. Further cooling is possible but reducing the temperature of the atoms would make the atoms more susceptible to heating during the following transportation stage.

After reducing the axial length of the atom cloud by evaporative cooling, we further transport the atoms to the surface. To do this, we use an extra beam (guide ODT), that intersects the center of the SIL and the atom cloud, providing a path to precisely move the atoms to the center of the SIL. After loading the atoms into this beam, the horizontal ODT is gradually removed (see figure \ref{figure1}(c)). The guide ODT is derived from a 532nm laser, with a power and beam waist of 2W and 20$\mu$m, respectively. The incident angle for this beam is also in the Brewster angle to prevent any unwanted reflections that would deteriorate the potential shape 

Finally, the vertical ODT is displaced by mechanically moving a mirror mounted over an air-bearing stage (Aerotech ABL10025) as shown in figure \ref{figure1}(d). The atoms move along the guide ODT to the center of the solid immersion lens, and the transportation of atoms is completed. The final number of atoms is $5 \times 10^5$ with a temperature of 2 $\mu$K, indicating that the temperature remained the same while 40\% of the atoms were lost during the second stage of transport. This loss is due to an inefficient transfer of the atoms from the horizontal ODT to the guide ODT, a small misalignment in the path of the vertical ODT during transport, and also  three-body losses. The final distance of the atoms from the surface of the SIL is 20$\mu$m. As the atomic cloud size is roughly spherical with a 20 $\mu$m diameter, further transportation of atoms will lead to loss of atoms due to van der Waals interaction with the surface of the SIL. Figure \ref{figure3}(a) shows the absorption images of atoms during transport. The probe beam is reflected by the surface of the SIL, generating two symmetrical images of the atomic cloud. By measuring the distance between the two clouds we can precisely estimate the distance from the surface.

\section{\label{sec:level3} Creation of a Bose-Einstein Condensate and Compression by optical accordion}

\begin{figure}[!ht]
\includegraphics{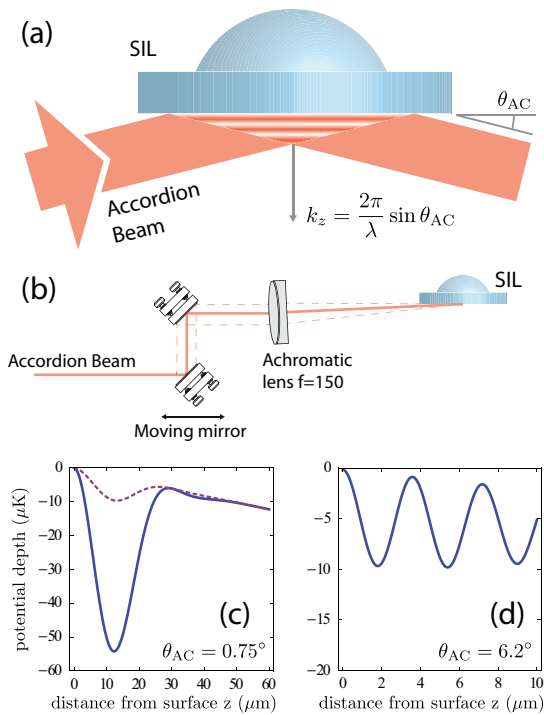}
\caption{Experimental setup of the optical accordion for compression of atoms. a) Reflecting the beam at the surface of the SIL produces a standing wave with a period which is function of the incident angle. b) The incident angle is manipulated using a lens and a moving mirror. c) d) Estimated potential in the vertical direction when the incident angle is 0.75 and 6.2 degrees, respectively. }
\label{figure2}
\end{figure}

To create a two-dimensional system of Yb atoms in an all-optical way, we used the ``optical accordion'' technique \cite{Li:08, PhysRevA.82.021604}, which consists of an one-dimensional optical lattice with manipulable periodicity. By reflecting a laser beam in a shallow angle into the SIL, a standing wave with a wavelength $\lambda_{AC} = \lambda / \sin \theta_{AC}$ dependent on the incident angle $\theta_{AC}$ is obtained, where $\lambda$ is the wavelength of the accordion beam (figure \ref{figure2}(a)). By manipulating the incident angle, we can compress the atomic cloud to create a two-dimensional system.

The apparatus used to change the incident angle of the accordion beam, consists of a lens and a moving mirror, as shown in figure \ref{figure2}(b). The lens is a 2 inch achromatic lens with a focus length of 200 mm. The focus point is aligned to the center of the SIL. When the mirror moves, the position of the beam axis displaces laterally, resulting in a change of the angle of the refracted beam. This setup allows us to change the incident angle by $\pm$ 14 degrees. The mirror is moved using an air-bearing mechanical stage to avoid mechanical vibrations (Aerotech ABL10050).

\begin{figure}[!ht]
\includegraphics{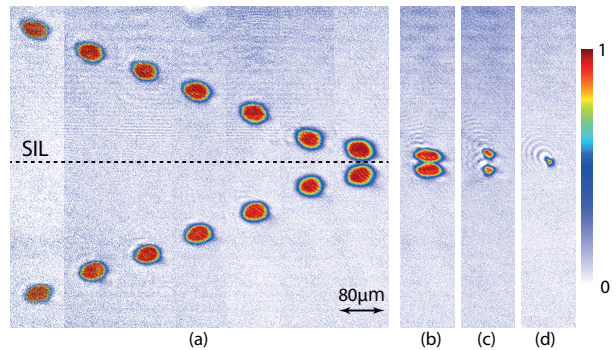}
\caption{(Zero time-of-flight) Absorption images of the transportation and compression process. The probe beam reflects in the surface of the SIL, producing two symmetrical images of the atom cloud. a) Transport of the atoms into a region 20$\mu$m below the surface. b) Transfer of atoms from the vertical ODT to the first layer of the optical accordion standing wave. c) BEC is created after forced evaporative cooling. d) After compression, atoms appear as a single spot in the absorption image due to the limited resolution of the imaging system.}
\label{figure3}
\end{figure}

The accordion beam is derived from a pulse laser centered at 780nm, with a repetition rate of 80MHz and a pulse duration of 1.3 ps. The coherence length of this pulse is approximately 500 $\mu$m, which is much shorter than the radius of the solid immersion lens ($R=5$ mm). Consequently, reflections inside the solid immersion lens will not coherently interfere with the incident beam, and a clean standing wave profile can be obtained. The compression procedure starts with an incident angle of 0.75 degrees, as shown in the solid curve of figure \ref{figure2}(c). At this angle, due to the Gaussian profile of the incident beam and the effect of gravity only one layer of standing wave forms. After adiabatically ramping up the intensity of the accordion beam in 100 ms, the vertical beam is gradually removed, and all the atoms are transferred into the first layer of the standing wave (figure \ref{figure3}(b)).

\begin{figure}[!ht]
\includegraphics{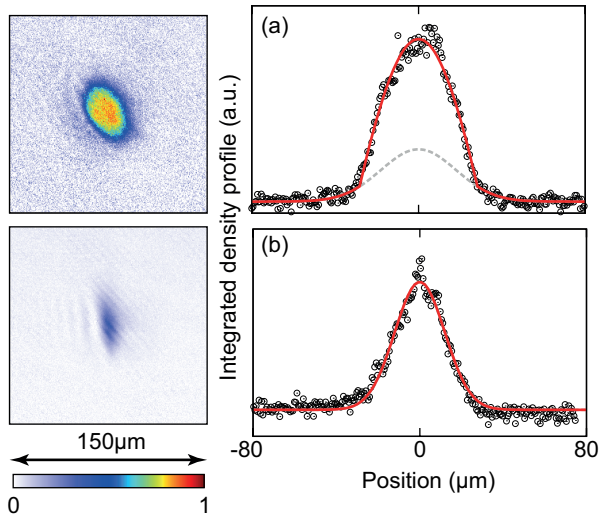}
\caption{Absorption images and density distributions of the atomic cloud after 5 ms of free expansion. a) BEC before the compression showing a typical anisotropic parabolic distribution along the long axis. The number of atoms in the BEC is $1.2 \times 10^4$ and the transition temperature is 170 nK. b) Optical accordion angle changes and atoms are compressed. Only 2000 atoms remain in the trap after a rapid loss due to three-body collisions. Atoms expand anisotropically, but the distribution along the vertical axis is Gaussian, showing that the cloud behaves as a two-dimensional system. }
\label{figure4}
\end{figure}

With the atoms loaded in the optical accordion, we perform a forced evaporative cooling of the atoms by exponentially decreasing the intensity of the accordion beam, as seen in figure \ref{figure2}(c) (dotted line), while keeping the intensity of the guide ODT constant. The initial number and phase-space density are $5 \times 10^5$ and $0.14$, respectively. After 3 seconds of evaporative cooling, a BEC of $1.2 \times 10^4$ atoms is obtained. The transition temperature, estimated using the thermal expansion of the bimodal distribution, was 170 nK. The absorption image after 5 ms of free expansion is shown in figure \ref{figure4}(a). Since the intensity of the guide ODT provides an strong confinement along the radial axis, the cloud expands rapidly in a direction almost perpendicular to the guide ODT, which is a clear signature of the condensate described by the Thomas-Fermi approximation.

Finally, the angle of the optical accordion is gradually increased to 6.2 degrees in 140 milliseconds, and the first layer is compressed as shown in figure \ref{figure2}(d). Only 2000 atoms remain in the trap after compression, positioned at 1.8 $\mu$m below the surface of the SIL. Before compression, the density of the condensate is $4 \times 10^{13}$ cm${}^{-3}$, resulting in a three-body loss rate of approximately 0.1 s$^{-1}$. During compression the density increases, leading to an increment of the three-body loss by more than two orders of magnitudes. As a result, atoms are rapidly lost due to three-body collisions. The atomic cloud after 5 ms of time-of-flight has also an anisotropic profile characteristic of condensates, with a density profile that is no longer parabolic (figure \ref{figure4}(b)) as would be expected in the Thomas-Fermi limit. After compression, the vertical trap frequency is 4.1 kHz, while the mean field energy is calculated to be $E_\text{mf} / 2 \pi \hbar =$ 640 Hz. As the mean field energy is considerably lower than the energy gap in the harmonic potential, all atoms reside in the ground state, and consequently, have a Gaussian momentum distribution. The resultant density distribution after compression, as shown in figure \ref{figure4}(b), is Gaussian, indicating that atoms are in a deep 2D regime, confined to the ground level of the harmonic potential in the vertical direction. The final shape of the atomic cloud is roughly a disc with a diameter of 5 $\mu$m and a thickness of 120 nm.

\section{\label{sec:level4}Conclusions}

We have demonstrated an all-optical method to transport Yb atoms to a distance of 20 $\mu$m 
below the surface of a solid immersion lens. The optical accordion technique was then utilized to 
create a Bose-Einstein condensate and compress the atoms into a thin layer of 1.8 $\mu$m below 
the surface of the lens. We were able to confirm that the atoms are in a deep 2D regime as a 
consequence of a strong confinement in the vertical direction created by the compressed optical 
accordion. The solid immersion lens, which increases the numerical aperture by a factor of the 
index of refraction of the lens, will provide a tool to observe and manipulate atoms with a high 
resolution microscope  \cite{nature09378, nature08482}. Our all-optical scheme will work with all 
atomic species regardless of the magnetic moment. 

We would like to thank Y. Takahashi, T. Mukaiyama, T. Kishimoto, S. Inouye, Y. Eto and K. Honda for their stimulating and fruitful discussions. This research is supported by the GCOE Program of MEXT through the Nanoscience and Quantum Physics Project of the Tokyo
Institute of Technology and the Cabinet Office, Government of Japan through its ``Funding Program for Next Generation World-Leading Researchers.''

\bibliography{apssamp}

\end{document}